\documentclass[prb,preprint]{revtex4}
\usepackage{graphicx}

\usepackage{subfigure}

\begin{document}

\title{Social applications of two-dimensional Ising models}

\author{D. Stauffer}

\affiliation{Institute for Theoretical Physics, Cologne University, 
D-50923 K\"oln, Euroland}

\begin{abstract}
I review three socio-economic models of economic opinions,
urban segregation, and language change and show that the well known 
two-dimensional Ising model gives about the same results in each case.
\end{abstract}

\maketitle

\section{Introduction}

Computer simulations of Ising models have been done for half a century, have been taught in courses on theoretical and computational physics, and are assumed to be known to the reader.\cite{landau} Here we discuss how simulations simulations can be applied to some problems of interest in the social sciences: urban segregation, language change, and economic opinions.

In the Ising model each site carries a spin which is up or down; neighboring spins ``like'' to be parallel, and the external field $h$ ``wants'' to orient the spins parallel to the field. At temperature $T$ the model is simulated with Boltzmann probabilities proportional to $\exp(-E/k_BT)$ by the Metropolis, Glauber, or the heat bath algorithms, or for constant magnetization $M$ using Kawasaki spin exchange.\cite{landau} The magnetization $M$ is the difference between the number of up and down spins and $E$ is the energy of the system. At low temperatures there is a spontaneous magnetization in zero field: Most of the spins are parallel to each other, either mostly up or mostly down. (For Kawasaki dynamics for which $M$ is fixed, the spins form two large domains.) At high temperatures the spins order into small clusters, without any long-range order. At very high temperatures in a field, the interaction between the spins becomes negligible. The spontaneous magnetization vanishes for $T \rightarrow T_c^-$, where $T_c \approx 2.27$ in units of the nearest-neighbor interaction energy. For $T$ slightly below $T_c$ in small lattices, $M$ switches irregularly between its positive and negative equilibrium values; for lower temperatures and a magnetization antiparallel to an external field, nucleation is possible and a small minority cluster parallel to the field can grow with a low probability to some critical size, and then grow rapidly to encompass the whole lattice. (In the literature this switching is called tunneling, but like nucleation it is not a quantum-mechanical effect.)

These basic algorithms and results can be taught in physics courses. We will see from the three following examples how these methods and results can be applied in social science.

\section{Business confidence}

In Germany important business managers are regularly asked how they judge the future economy, and their opinions are summarized and published in the Ifo index in the form of positive, negative, and neutral opinions, as well as an overall average.\cite{ifo}
Hohnisch\cite{hohnisch} noted that there is a tendency for either a strongly pessimistic or strongly optimistic opinion to form, similar to the Ising simulations shown in Fig.~1. Thus psychological interactions (``herding'') between the managers exist besides hard economic facts.

\begin{figure}[t]
\begin{center}
\includegraphics[angle=-90,scale=0.5]{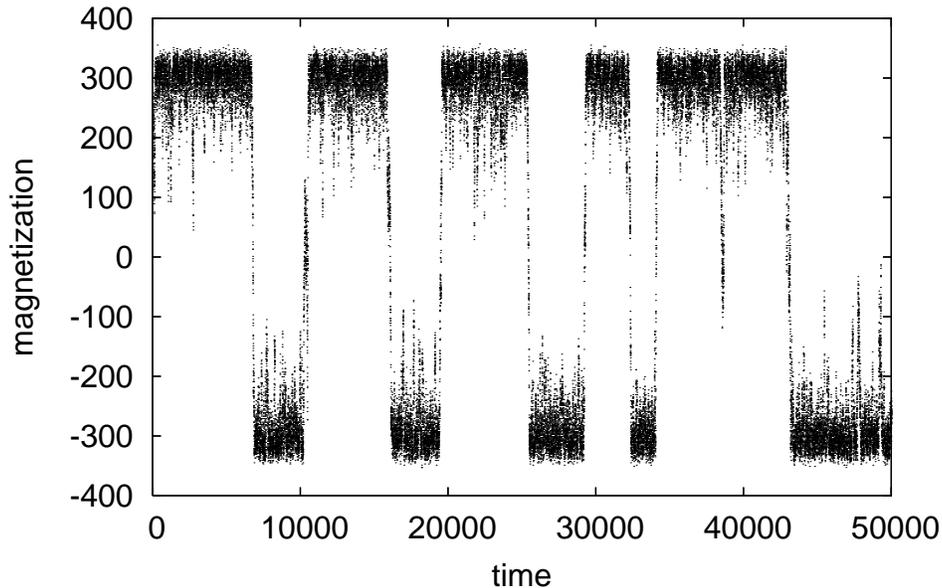}
\end{center}
\caption{Switching of the Ising model on a a $19 \times 19$ square lattice at $T=2.2$ from one sign of the spontaneous magnetization to the opposite. This change can be interpreted as swings in business confidence\cite{hohnisch} or as changes in a language feature.\cite{nettle} Time is measured in iterations; each site is updated once in an iteration. Adapted from Ref.~\onlinecite{langssw}.}
\end{figure}

Apparently, the polled experts influence each other, like ferromagnetic spins. The high-tech bubble of the U.S.\ stock markets, which ended in spring 2000, might have been another example of psychological impact on economics: First numerous people thought that everything having to do with computers would make big profits, and then doubts spread. In 1971 Weidlich published a more complicated method of treating such binary opinions.\cite{weidlich}

Reference~\onlinecite{hohnisch} was followed by a study of the influence of temporally fluctuating external fields; too much such noise destroys the spontaneous magnetization.\cite{isshock} If there is good (bad) news, people may evolve to form an optimistic majority, but if good news too quickly changes into bad news and back, then people just ignore it and adopt random attitudes. We might think of curriculum reform in societies were the government sets the rules and changes them before the first students under the old rules have finished their studies.\cite{baumert}

\section{Schelling's segregation}

In 1971 the 2005 Nobel economist Thomas Schelling published an Ising-like model\cite{schelling} to explain racial segregation in U.S.\ cities due to the personal preferences of the residents, without any outside pressures (for example, different housing costs and discrimination). As we will discuss, his model roughly corresponds to an Ising model at $T = 0$. It took 35 years before it was clearly realized that in contrast to reality (like the old Harlem in New York City) no large domains are formed in Schelling's model,\cite{kirman} even though Schelling's paper had become a classic in urban dynamics.\cite{fossett} Schelling's model was corrected, for example, by Jones,\cite{jones} who introduced random dynamics: A small fraction of people move away and are replaced by people who are happy in their new residence. As a result large domains are formed as desired. This paper was widely ignored. The simple Ising model at low $T$ is a much earlier example of the self-organization of such domains.

Physicists were not aware of Ref.~\onlinecite{schelling} for nearly three decades,\cite{lls} but then simulated Ising and Potts models, with two and more groups for increasing values of $T$, as suggested by Weidlich. The temperature can be interpreted as a measure of the tolerance toward people of different groups. If $T$ increases fast enough, large domains can be avoided,\cite{ortmanns} with both two and more than two groups. (With more than two groups and conserved group membership as in Kawasaki kinetics, there is one domain for each group in low temperature equilibrium.) 

At fixed $T$, the various complications by which the Schelling model differs from the simple Ising model are hardly relevant: usually $T = 0$ gives small clusters and $0 < T < T_c$ gives large domains.\cite{solomon} This lack of importance is also true for various versions that are intermediate between the Ising and the Schelling model.

More interesting than these intermediate cases is the feedback where each site has its own value of $T$. This value increases (decreases) if all neighbor spins have the same (opposite) orientation as the spin of interest; there also is a general tendency of $T$ to decrease, that is, for a person to forget his/her own tolerance. (Some drastic event shocks people into more tolerant behavior, and then they slowly forget.\cite{kantz}.) In this way a stationary average ${<}T{>}$ self-organizes, with or without spontaneous magnetization depending on the forgetting rate. The usual way of assuming from the outset a social temperature $T$ is thus avoided by this self-organization, depending on the forgetting rate. Figure~2 shows how a change from a forgetting rate of 0.2 percent to 0.3 percent per iteration lowers only slightly the mean temperature, but creates a large spontaneous magnetization (which is analogous to urban segregation and large ghettos). This model suggests that life-long reinforcing of tolerance is needed to avoid ghetto formation.

\begin{figure}[hbt]
\begin{center}
\includegraphics[angle=-90,scale=0.5]{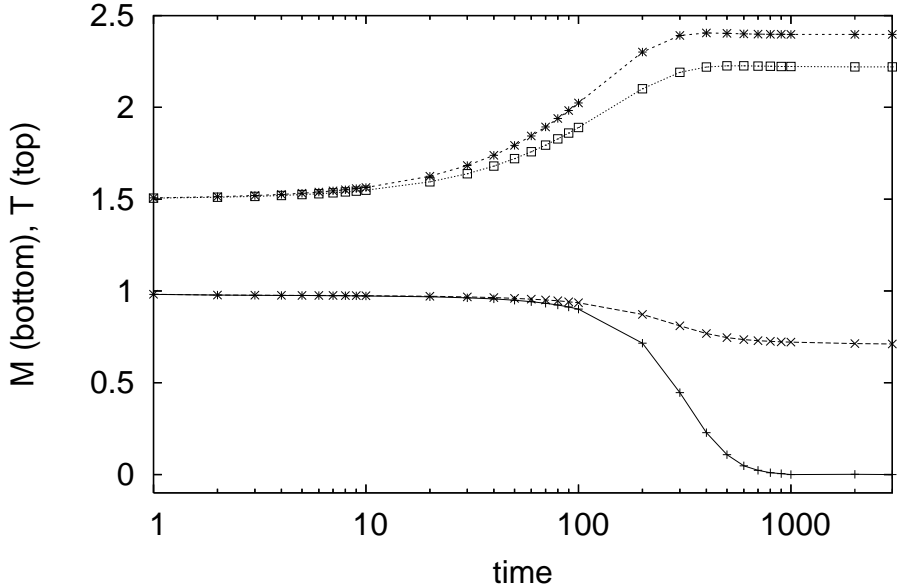}
\end{center}
\caption{A $7001 \times 7001$ Ising model on a square lattice with a self-organizing local temperature due to feedback with the neighborhood. The two top curves give the average temperature, the two lower curves give the magnetization for temperature decreases (``forgetting'') of 0.2\% (+,stars) and 0.3\% (x,squares) per iteration. Adapted from Ref.~\onlinecite{solomon}.}
\end{figure}

\section{Nettle's language change}

Nettle\cite{nettle} simulated how one language (or language feature) which is spoken by everybody can be replaced by another language (feature) without any outside force or bias. His Fig.~1 is reminiscent of our Fig.~1: irregular switching between positive and negative spontaneous magnetizations. He found that the rate at which the majority switches language decays if the population becomes larger; in the Ising model the answer is already known: the rate decays exponentially with increasing linear dimension of the lattice.\cite{trapp} In the simulation of Ref.~\onlinecite{nettle} not everybody was equal: some speakers were more equal than others, as in social impact theory.\cite{latane} This complication made the model more realistic, but is not needed to get language change.

If we consider a very large lattice, the switching of Fig.~1 become exponentially rare. If we introduce a bias in favor of the new language (feature), the bias corresponds to an external field, and the previously mentioned nucleation process occurs, as has been well studied for Ising models and for languages.\cite{nettle,wang}

Later work\cite{langchange} replaced a single Ising spin by 8 language features each of which can have 5 values as in the Potts model. The size dependence of the language change rate depends weakly on population size if people learn language features only from their neighbors, and depends strongly on the population size if people can learn from all other members of the population. More details are given in Refs.~\onlinecite{langchange} and \onlinecite{langssw}. Because different simulations give different results for the size dependence of language change, much more work is needed to clarify that question of whether languages spoken by many people change more slowly than than those spoken by less people. In comparison, the Ising model has been simulated for nearly half a century. And also empirical work on change in real in contrast to simulated languages give conflicting results.\cite{nettle,langchange}

\section{Discussion}

The three examples have shown how the standard Ising model can be applied to socioeconomic questions. In two cases the original authors\cite{schelling,nettle} were not aware of the Ising model and made their models more complicated than necessary to answer their questions. Of course, these complications may make the model more realistic and were simulated even in the first example for which the Ising model was known to the authors.\cite{hohnisch} Had the social scientists learned about Ising models, or had this author read more carefully the book\cite{lls} in which Schelling was cited by several physicists, progress could have been faster.

How can we use this material for teaching in physics classes? While teaching electricity and magnetism, I introduce ferromagnetism and the Ising model in connection with Maxwell's equations in matter, where the magnetization gives the difference between the magnetic $B$ and $H$ fields and where the concept of a spontaneous magnetization can be introduced. Then the above examples can be presented. I used Schelling's model when ethnic segregation was much in the news in Germany in the author's home town.

\appendix*

\section{Further details of the simulations}

\begin{figure}[hbt]
\begin{center}
\includegraphics[angle=-90,scale=0.35]{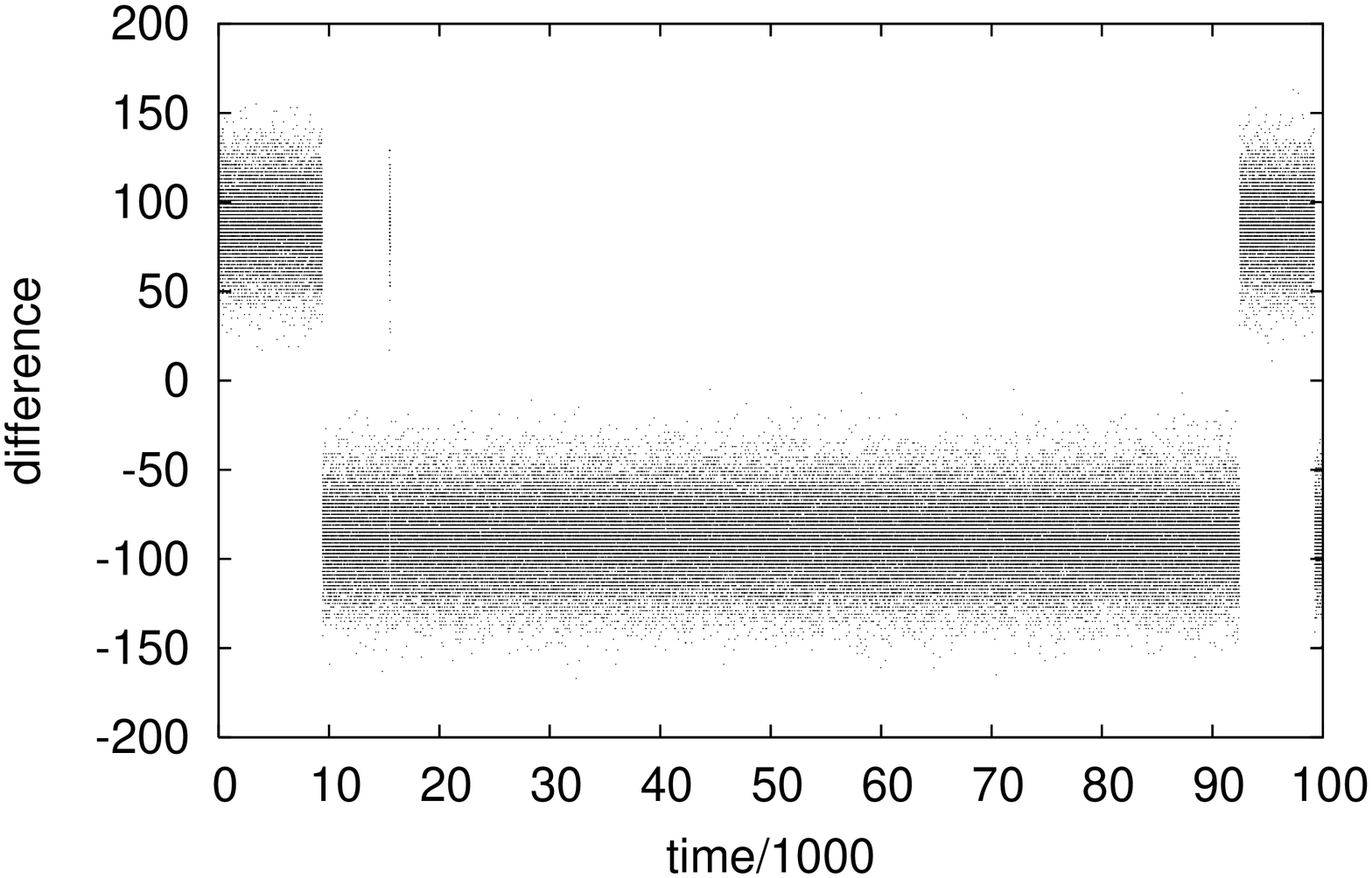}
\includegraphics[angle=-90,scale=0.35]{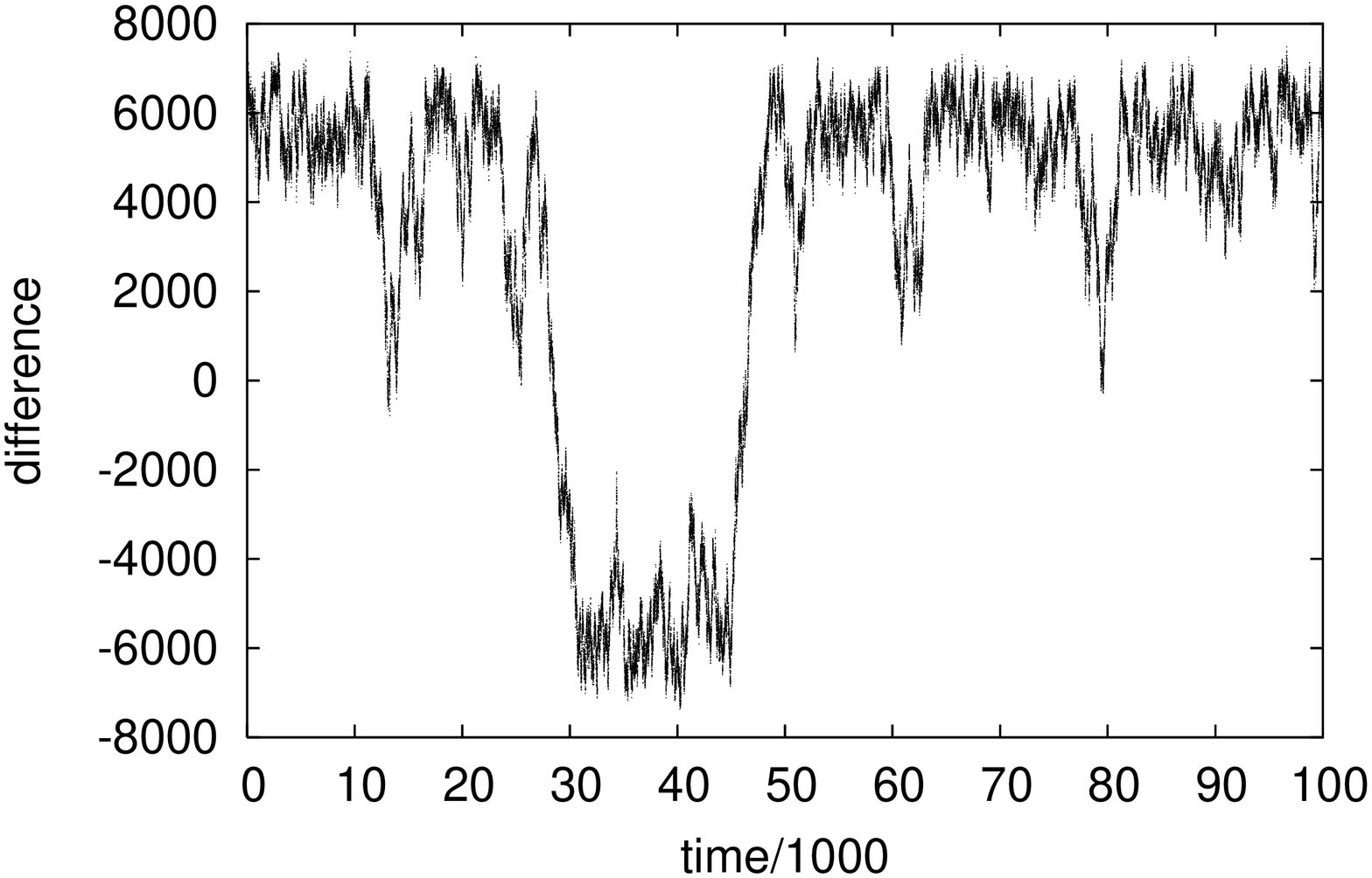}
%\subfigure[\ ]{\scalebox{0.7}{\includegraphics[angle=-90,scale=0.39]{isinggould3a.eps}}}
%\subfigure[]{\scalebox{0.7}{\includegraphics[angle=-90,scale=0.39]{isinggould3b.eps}}}
\end{center}
\caption{Nettle-type simulation of language switching without bias and without influential and less influential speakers. (a) refers to everybody interacting with everybody (401 speakers, 39 \% noise) and (b) has the usual nearest-neighbor interaction of a $101 \times 101$ square lattice at 7.7 \% noise.\cite{langchange}}
\end{figure}

{\it Hohnisch}.
The simulations in Ref.~\onlinecite{hohnisch} are more complicated than the Ising model and allow also for a neutral opinion, as in the Blume-Capel and Potts models.\cite{BC,wu} In the Blume-Capel model there are three choices $+1$, 0, and $-1$ for each spin $S_i$, and a term proportional to $S_i^2$ is added to the usual interaction energy $\sum_{i,k}S_iS_k$. This additional term controls the fraction of neutral (zero) spins. In the Potts model, each spin has an integer value between 1 and $q$, and the interaction energy of two spins has a low value if the two spins are the same, and a high value if they are different. The special case of $q=2$ corresponds to the usual Ising model. For a fluctuating external field\cite{hohnisch} $B = \pm b$ applied to a fraction $p$ of the spins, there is a spontaneous magnetization if $b$ is less than some critical value $b_c$ but not above it. The value of $b_c$ decreases if $T \rightarrow T_c^-$ and/or if the influenced fraction $p$ increases. These fluctuating fields can be interpreted as external news that is good or bad.

{\it Schelling}.
Schelling\cite{schelling} defines the neighborhood as the eight nearest and next-nearest neighbors on a square lattice. Each site is either empty, or occupied with someone from one of the two groups of people. People are defined as happy if at least half their neighbors are from their own group, and otherwise are unhappy. Unhappy people move to the closest empty residence where they are happy. Many variants were also studied by Schelling. Schelling's and Nettle's model may
be regarded as (stochastic) cellular automata, but with sequential instead
of the usual simultaneous updating.

{\it Nettle}. Nettle \cite{nettle} assumed that each site follows the majority of the neighbors, except that with some noise probability it does the opposite. Even for the case of everybody influencing everybody, and everybody being equally important, switching is possible, see Fig.~3(a). This simplification corresponds to a majority rule or voter model\cite{liggett} with noise.

In Fig.~3(a) we assumed that the spin $S_i = \pm 1$ to change as $S_i = {\rm sign}(\sum_k S_k)$, where $k$ runs over all sites and the sign function is replaced by a random choice if the spin sum gives zero. Then with some probability, taken here to be 39 percent, the spin is reversed (independently on whether it was reversed before). Similar results were obtained in Fig.3(b) on a $101 \times 101$ lattice with only 7.7 percent noise and influence only from the four nearest neighbors. 

\begin{acknowledgements}
I thank S.\ Wichmann, M.\ Hohnisch, E.\ Holman, K.\ M\"uller, S.\ Pittnauer, C.\
Schulze, and S.\ Solomon for research collaborations in this field.
\end{acknowledgements}

\end{document}